\documentstyle[11pt,psfig,twoside]{article}

\begin{document}

\newcommand{\dd}{\,{\rm d}}
\newcommand{\ie}{{\it i.e.},\,}
\newcommand{\etal}{{\it et al.\ }}
\newcommand{\eg}{{\it e.g.},\,}
\newcommand{\cf}{{\it cf.\ }}
\newcommand{\vs}{{\it vs.\ }}
\newcommand{\zdot}{\makebox[0pt][l]{.}}
\newcommand{\up}[1]{\ifmmode^{\rm #1}\else$^{\rm #1}$\fi}
\newcommand{\dn}[1]{\ifmmode_{\rm #1}\else$_{\rm #1}$\fi}
\newcommand{\upd}{\up{d}}
\newcommand{\uph}{\up{h}}
\newcommand{\upm}{\up{m}}
\newcommand{\ups}{\up{s}}
\newcommand{\arcd}{\ifmmode^{\circ}\else$^{\circ}$\fi}
\newcommand{\arcm}{\ifmmode{'}\else$'$\fi}
\newcommand{\arcs}{\ifmmode{''}\else$''$\fi}
\newcommand{\MS}{{\rm M}\ifmmode_{\odot}\else$_{\odot}$\fi}
\newcommand{\RS}{{\rm R}\ifmmode_{\odot}\else$_{\odot}$\fi}
\newcommand{\LS}{{\rm L}\ifmmode_{\odot}\else$_{\odot}$\fi}

\newcommand{\Abstract}[2]{{\footnotesize\begin{center}ABSTRACT\end{center}
\vspace{1mm}\par#1\par
\noindent
{~}{\it #2}}}

\newcommand{\TabCap}[2]{\begin{center}\parbox[t]{#1}{\begin{center}
  \small {\spaceskip 2pt plus 1pt minus 1pt T a b l e}
  \refstepcounter{table}\thetable \\[2mm]
  \footnotesize #2 \end{center}}\end{center}}

\newcommand{\TableSep}[2]{\begin{table}[p]\vspace{#1}
\TabCap{#2}\end{table}}

\newcommand{\FigCap}[1]{\footnotesize\par\noindent Fig.\  %
  \refstepcounter{figure}\thefigure. #1\par}

\newcommand{\TableFont}{\footnotesize}
\newcommand{\TableFontIt}{\ttit}
\newcommand{\SetTableFont}[1]{\renewcommand{\TableFont}{#1}}

\newcommand{\MakeTable}[4]{\begin{table}[htb]\TabCap{#2}{#3}
  \begin{center} \TableFont \begin{tabular}{#1} #4 
  \end{tabular}\end{center}\end{table}}

\newcommand{\MakeTableSep}[4]{\begin{table}[p]\TabCap{#2}{#3}
  \begin{center} \TableFont \begin{tabular}{#1} #4 
  \end{tabular}\end{center}\end{table}}

\newenvironment{references}%
{
\footnotesize \frenchspacing
\renewcommand{\thesection}{}
\renewcommand{\in}{{\rm in }}
\renewcommand{\AA}{Astron.\ Astrophys.}
\newcommand{\AAS}{Astron.~Astrophys.~Suppl.~Ser.}
\newcommand{\ApJ}{Astrophys.\ J.}
\newcommand{\ApJS}{Astrophys.\ J.~Suppl.~Ser.}
\newcommand{\ApJL}{Astrophys.\ J.~Letters}
\newcommand{\AJ}{Astron.\ J.}
\newcommand{\IBVS}{IBVS}
\newcommand{\PASP}{P.A.S.P.}
\newcommand{\Acta}{Acta Astron.}
\newcommand{\MNRAS}{MNRAS}
\renewcommand{\and}{{\rm and }}
\section{{\rm REFERENCES}}
\sloppy \hyphenpenalty10000
\begin{list}{}{\leftmargin1cm\listparindent-1cm
\itemindent\listparindent\parsep0pt\itemsep0pt}}%
{\end{list}\vspace{2mm}}

\def\TYLDA{~}
\newlength{\DW}
\settowidth{\DW}{0}
\newcommand{\dw}{\hspace{\DW}}

\newcommand{\refitem}[5]{\item[]{#1} #2%
\def\REFARG{#3}\ifx\REFARG\TYLDA\else, {\it#3}\fi
\def\REFARG{#4}\ifx\REFARG\TYLDA\else, {\bf#4}\fi
\def\REFARG{#5}\ifx\REFARG\TYLDA\else, {#5}\fi.}

\newcommand{\Section}[1]{\section{#1}}
\newcommand{\Subsection}[1]{\subsection{#1}}
\newcommand{\Acknow}[1]{\par\vspace{5mm}{\bf Acknowledgments.} #1}
\pagestyle{myheadings}

\def\thefootnote{\fnsymbol{footnote}}

\begin{center}
{\Large\bf Period Changes of the SMC Cepheids from the Harvard,
OGLE and ASAS Data}
\vskip1cm
{\bf
P~a~w~e~\l\ ~~P~i~e~t~r~u~k~o~w~i~c~z}
\vskip3mm
{Warsaw University Observatory, Al. Ujazdowskie 4, 00-478 Warszawa, Poland\\
e-mail: pietruk@astrouw.edu.pl}
\end{center}

\Abstract{
Comparison of the old observations of Cepheids in the Small Magellanic
Cloud from the Harvard data archive, with the recent OGLE and ASAS
observations allows an estimate of their period changes. All of
matched 557 Cepheids are still pulsating in the same mode.
One of the Harvard Cepheid, HV 11289, has been tentatively matched
to a star which is now apparently constant.
Cepheids with ${\rm \log ~ P > 0.8 }$
show significant period changes, positive as well as negative.
We found that for many stars these changes are significantly smaller than
predicted by recent model calculations. Unfortunately, there are no models
available for Cepheids with periods longer than approximatelly
80 days, while there are observed Cepheids with
periods up to 210 days.
}{Stars: evolution -- Cepheids -- Magellanic Clouds}

\Section{Introduction}

Classical Cepheids are the most popular standard candles for
extragalactic distance estimates. They are also useful
for testing models of stellar interior and evolution.

Cepheids are massive Population I stars crossing the instability strip
in the Hertzsprung--Russell diagram at the effective temperature 
${\rm \log T_{eff} \approx 3.8}$. Certainly most of them are
undergoing core helium burning. There is also a possibility that 
a small fraction of observed Cepheid may be crossing the strip when
the star is in the Hertzsprung gap and evolves on the thermal time scale.
This crossing the instability strip is termed crossing I.  
The two that follows, termed crossings II and III,
occur during helium burning and are generally much slower.

While a star crosses the instability strip its pulsation period
changes. Even for massive objects the evolutionary period changes
are very slow and a long time interval is needed to detect them.
Some Cepheids in our Galaxy has been observed for about 200 years,
e.g. $\delta$ Cep, the prototype of this group, was discovered
by John Goodricke in 1784.

For many stars significant period
changes were detected (Berdnikov and Ignatova 2000).
Recently there were also published extensive observations of some 
Galactic Cepheids exhibiting large period and/or amplitude changes,
which are unlikely of evolutionary origin,
e.g. Polaris (Kamper and Fernie 1998, Evans et al. 2002) or Y Oph
(Fernie 1995). Also Turner (1998) presented data on period
changes of 137 northern hemisphere Cepheids. Earlier a quantitative
relation between the observed changes and those predicted
by the evolutionary models was investigated by Hofmeister (1967).
Saitou (1989) tried to find effects of metal abundence on the
evolutionary period changes and concluded that there is a marginally
dependence. However, this reasoning was based on 37 stars only
and the influence of errors was not taken into account.
Recently Macri, Sasselov and Stanek (2001)
reported on a dramatic change in the light curve of a Cepheid
discovered by Hubble in M33. They suggest that the star
stopped pulsating.

In the Magellanic Clouds almost four thousand Cepheids are known.
The first large database, containing periods, moments of maxima
and magnitudes of the SMC Cepheids, was published
by Payne-Gaposchkin and Gaposchkin (1966, hereafter PG\&G).
It is a result of long time photographic survey conducted in the Harvard
observatories in the years 1888 -- 1962. Later Deasy and Wayman
(1985) found that about 40 percent of a sample of 115 stars showed
period variations, apparently too rapid to be explained
with the evolutionary models. In the late 1990's a rich
observational material  for the Magellanic Cloud Cepheids
was obtained by several groups searching for gravitational microlensing.
In this paper we determine period changes in the SMC Cepheids
comparing the data published by PG\&G with the results of
two recent projects: OGLE (the Optical Gravitational
Lensing Experiment, Udalski et al. 1997), and ASAS (the All Sky
Automated Survey, Pojma\'nski 2000). We also compare the observed
period changes with the predictions of the recent stellar
evolutionary models. 

\Section{Observational Data}

PG\&G dataset contains 1201 periodic variable stars,
mostly classified as classical Cepheids. The remaining
variables of the group are foreground RR Lyr or W Vir
stars. Each Cepheid has its HV (Harvard Variable) number.
The positions are given in rectangular coordinates, probably as
defined on a reference photographic plate by Leavitt (1906).
This created some problems with finding the most precise equatorial
coordinates. Moreover, the analysis was complicated by the fact
that subsequent Harvard observers used different instruments.

The PG\&G database gives also the moments of maxima corresponding
to the best observed epochs. For variables with
detected period variations there is more information, like
a suggested period in some epochs and occasionally
parabolic elements for O-C diagrams.
Contemporary data for the fainter Cepheids are taken from the OGLE-II 
(Udalski et al. 1999b) and for the brightest stars --
from the ASAS (Pojma\'nski 2002, in preparation) projects.

The OGLE and Harvard databases were matched using 2000.0 coordinates.
For each Cepheid from Harvard list, which should 
lie in one of 11 OGLE fields, we looked for an OGLE Cepheid in 
a square 80 $\times$ 80 arcsec. If there were more than one star in the
square, we chose that with a very close period. In this we identified
534 Cepheids. Six more were found in the new OGLE II database
of variable stars (\.Zebru\'n et al. 2001),
obtained with a new data reduction software, called Diffrence Image
Analysis or ISIS (cf. Alard and Lupton 1998, Alard 2000), Wo\'zniak 2000).

However, we could not match 14 Cepheids. Four of them (HV 821, HV 824
HV 829, HV 1956) are too bright for the OGLE camera (their images are
saturated). Fortunately they were easily identified in the ASAS data.
Two Cepheids, HV 1933 and HV 1726, are located close to the
border of the OGLE area in the SMC, could not be found probably
due to small errors of their coordinates in PG\&G catalog.
Two other stars, HV 1959 and HV 11174, have periods very close to
a multiple of a day; such variables were rejected by the OGLE.
HV 1369 is likely not to be a Cepheid, because it is located outside
the area covered by the Cepheids in the Fourier decomposition parameters
$vs. \log P$ diagrams presented by Udalski et al. (1999b, Fig. 3).
For the remaining 5 Cepheids no pulsating OGLE stars was found,
either in the list of single mode or double mode Cepheids prepared by
Udalski et al. (1999a). Four stars: HV 1353, HV 1714,
HV 1796, HV 11483, do not have even a constant counterpart
within their expected ranges of magnitudes and
a radius of 30 arcsec. HV 11289 (with  $ P=0.788$ d) may have
its constant OGLE counterpart. If this identification is correct
then a star pulsating with an amplitude of 0.7 mag stopped its
pulsations in just 50 years. Definitive identification should be
possible from the comparison of the original Harvard
plates with public domain OGLE images.
One should remember about possible typing errors in coordinates
given by PG\&G though they were checked in original papers
(Leavitt 1906, Nail 1942). 

In addition to OGLE variables the ASAS provided data for seventeen
brightest Cepheids. Therefore, we disposed
a total of 557 variables for further analysis.
This sample for the SMC is larger than a sample presented
for the LMC Cepheids by Pietrukowicz (2001, hereafter Paper I)
which had 378 stars.

To be sure that stars were matched correctly we compared the
magnitudes (Fig. 1) and coordinates (Fig. 2) obtained from
the Harvard catalog and from the OGLE or ASAS catalogs. Although
there are large discrepancies in one of the coordinates for several
stars, we did not reject any of them. We note that among 557 Cepheids
42 are the first overtone pulsators. Cross-correlations of each
variable and its parameters are available on the Internet at
\begin{center} 
ftp://ftp.astrouw.edu.pl:/pub/pietruk/cephSMC.tab
\end{center}

\Section{Evolutionary Models of Cepheids}

The most important properties of all evolutionary models were
described in Paper I. A recent theoretical
survey of Cepheids' characteristics for a number of
evolutionary models was published by ABHA (Alibert,
Baraffe, Hauschildt, Allard 1999). It contains parameters of 
stars at the blue and red edges of the instability strip for models in 
the ZAMS mass range ${\rm 3 - 12 M_\odot }$ with chemical 
composition ${\rm Z=0.004, Y=0.25 }$, representative for the SMC.

Another theoretical set of models were recently published by Bono et al.
(2000). They adopted the same metallicity as ABHA, but two different
helium contents: $Y=0.23$ and $Y=0.27$. Using a linear nonadiabatic
pulsation code, kindly provided by Dr. W. Dziembowski,
we calculated values of the period changes for the Bono et al. (2000)
models in about twenty points of time for each crossing through
the instability strip.

For the ABHA models we have values of the pulsation periods
$P_0$ and $P_1$ in two moments of time $t_0$ and $t_1$ respectively
(at the strip edges). We define the theoretical
rate of period change as
$$
r_{th} \equiv \frac{\Delta P}{\Delta t}\frac{1}{{P}^2}
=\frac{P_1-P_0}{t_1-t_0}\frac{1}{{P}^2}
\eqno(1)
$$
The scaling is chosen so that all model results and observational
points can be cleary displayed in the figures which follow.

\Section{Comparison with Observations}

We calculate the rate of the observed period change using the equation 
$$
r_{obs} \equiv \frac{\Delta P}{\Delta t}\frac{1}{{P_1}^2}
=\frac{P_1-P_0}{t_1-t_0}\frac{1}{{P_1}^2}
\eqno(2)
$$
where $P_0$ is the old (Harvard) period at the moment of Cepheid light curve
maximum $t_0$, and $P_1$ is the new (OGLE or ASAS) period at the moment of 
maximum $t_1$. We estimate the uncertainty of the rate of period change 
using the relation:
$$
\sigma_{obs} \approx \frac{\sigma _{P_1}}{t_1-t_0}\frac{1}{{P_1}^2}
\eqno(3)
$$ 
where $\sigma _{P_1}$ is the estimated error of the period.
Both $P_1$ and $\sigma _{P_1}$ were returned by a program
by Schwarzenberg-Czerny (1996).
Unfortunately, the error estimates of the Harvard periods,
$\sigma _{P_0}$, were not given by PG\&G. Therefore, $\sigma_{obs}$
is the lower limit of the observational error of the rate.
However, the periods determined from Harvard data are generally of high 
accuracy, as they are based on the observations covering several decades.
Hence, the real $\sigma_{obs}$ is not likely to be much larger than the
estimate given by Eq. (3). We neglected the contribution of
${\rm t_0 }$ and ${\rm t_1 }$ uncertainties to the error budget.

Just as it was in the case of Cepheids in the LMC (Paper I), Cepheids
with the longest periods, ${\rm log ~ P > 0.8}$ have significant period
changes. However we note that there are several stars with
${\rm log ~ P < 0.8}$, which also have measurable changes.
A comparison between the rates of period change for the
fundamental mode Cepheids and ABHA models is presented in Fig. 3.
The theoretical predictions for the three instability crossings
are clearly separated. None of the star appears to undergo the first
crossing, which corresponds to the evolution on the thermal time scale.
For Cepheids with long periods the changes are much smaller
than predicted by models with metallicity ${\rm Z=0.004}$, expected
for SMC stars. However, HV 829 appears to be changing its period
at a rate expected from extrapolating model predictions.

Fig. 4 and Fig. 5 show two comparisons between two sets of
theoretical models from Bono et al. (2000)
for the same metallicity, but for two helium contents,
and the observations determined from the Harvard,
OGLE and ASAS data. It is important to notice that the highest
masses are ${\rm 11 M_\odot}$ and ${\rm 12 M_\odot}$,
for ${\rm Y=0.23}$ and ${\rm Y=0.27}$ respectively.
For a given period the range of period changes
in crossing III is slightly larger for ${\rm Y=0.27}$.
For most Cepheids with ${\rm log ~ P > 0.8}$ the observed
period changes are smaller than indicated by both sets of models.
Many ASAS stars seem to have large changes, but their
errors are large as a consequence of a small number of epochs.
However we are confident that HV 829 decreases its period relatively
fast. PG\&G already found a secular period change of this star
and they estimated the moments of maximum brightness as
$$ 
M = 11508.062 + 89.92 E - 0.006 E^2
\eqno(4)
$$
where E is the epoch number. Our rate of period change,
${\rm \dot P/{P^2} = (-19.0 \pm 0.8) }$ 
${\rm 10^{-9} d^{-2} }$,
is consistent with those calculations at one sigma level.
Therefore the change rate is likely to be constant. 

Two Cepheids in the SMC have extremely long periods:
126.3 and 212.5 days. Unfortunately we cannot compare their
characteristics with theoretical predictions, as suitable models do not
exist.

Fig. 6 displays a comparison between the period changes
determined from the models given by Bono et al. (2000)
and the observed period changes of the first overtone Cepheids.
Generally the models agree with observations. But it is
clear that the observed changes are hardly significant.
However, some stars apear to evolve faster than it is predicted.
HV 12937 has a strange rate of period change. There may be
an error in the period value of this star in PG\&G database,
although its old and new coordinates are in good agreement.
Other three variables (HV 1384, HV 11167, HV 11196)
change their periods comparably fast to the very well-known Polaris,
which in our units would have 
${\rm \dot P_{fo}/{P_{fo}^2} = (6.28 \pm 0.20) }$
${\rm 10^{-9} d^{-2} }$
with ${\rm log ~ P_{fo} = 0.599 }$ (based on Berdnikov and Ignatova 2000).

\Section{Discussion}

Our analysis led to several interesting conclusions. We identified
only one Cepheid: HV 11289, which may have stopped pulsating
between the Harvard and OGLE and ASAS epochs. This has to be
verified especially because of possible wrong coordinates
given by PG\&G. We do not expect to find many such stars.
The evolutionary models for ${\rm Z=0.004}$ predict the time
of the crossing in the loop phase as $\sim 10^5$ for
${\rm P=10}$ days, and $\sim 10^4$ for ${\rm P=30}$ days.
Hence the probability of leaving the instability strip in one century by
a long period Cepheid is approximately $5 \times 10^{-3}$.
The probability is smaller for Cepheids with shorter periods,
i.e. less massive stars. Similar estimates are valid for
the probability of mode switching.
Except for HV 11289 all matched Harvard Cepheids are still
pulsating in the same mode, which is not surprising in view
of our estimate.

We found that no star is undergoing the first crossing,
which represents a rapid evolution on the thermal time scale.
Theory predicts the first crossing time to be a few tens times shorter than
times for crossings II or III. Therefore, we expected to find
several Cepheids with period changes corresponding to the crossing
of Hertzsprung gap. Only one star, a first overtone pulsator
HV 12937, has a very large rate of period change (cf. Fig. 6), but it
is negative, i.e. it cannot correspond to the first crossing.
This situation is similar to that presented in Paper I for the
LMC sample.

We found that Cepheids with ${\rm log ~ P > 0.8}$ have significant
period changes, but for many of them one cannot decide which
crossing (II or III) they are undergoing.
Generally the changes are small and a few times slower than
the lowest values for crossings on the nuclear time scale
calculated from models given by Bono et al. (2000), as presented
in Fig. 7. Moreover, we found that their calculations
for ${\rm Z=0.004}$
and ${\rm Y=0.23}$ predict that a star with the mass ${\rm 12 M_\odot}$
makes a small loop but it does not reach the instability strip
during the core helium burning. The discrepancy between the ABHA
models and the observations for long period Cepheids is even larger.
This theoretical survey also predicts too few long period Cepheids.
Meanwhile, the observations confirm that there are Cepheids
with periods up to 210 days in the Small Magellanic Cloud.
Therefore we conclude that the predictions for massive stars,
i.e. long period Cepheids, given by both sets
of models cannot be right. For the first overtone Cepheids the
expected and the observed rates of period changes are of the same order
of magnitude, but in this case the accuracy is insufficient for a firm
conclusion.

There is a good prospect for improvement in observational
constraints on the rate of period changes.
In near future (5--10 years) one will be able to phase
them and achieve much better estimates
of the period changes and their errors. Our present analysis
relates only to about a half of the Cepheids listed in Harvard archives.
New observations covering the entire Magellanic Clouds regions will
extend the catalog. A comparison between predicted and observed
period changes for hundreds of Cepheids in our Galaxy would be
also interesting and would help to refine the theory of stellar
structure and evolution.

\Acknow{

I would like to thank Dr. G. Pojma\'nski for providing 
photometric data on the brightest SMC Cepheids before publishing,
Dr. A. Schwarzenberg-Czerny for software useful to search precise
periods and Dr. P. Moskalik for providing the list of previous papers
on the Cepheid period changes. I also thank Dr. S. Cassisi
for making available the set of evolutionary tracks calculated by
Bono et al. (2000). I am greatful to Dr. W. Dziembowski for
providing the pulsation code and important discussions.
I would like to thank Dr. K. Z. Stanek for useful comments
and Dr. B. Paczy\'nski for remarks and an insight to one
of the original papers containing Harvard data.
I wish to thank I. Soszy\'nski and K. \.Zebru\'n, OGLE team
members, for valuable explanations and helpful software.
Support by the BW grant to Warsaw University Observatory
is acknowledged.
}

\begin{figure}[htb]
\hglue-0.5cm\psfig{figure=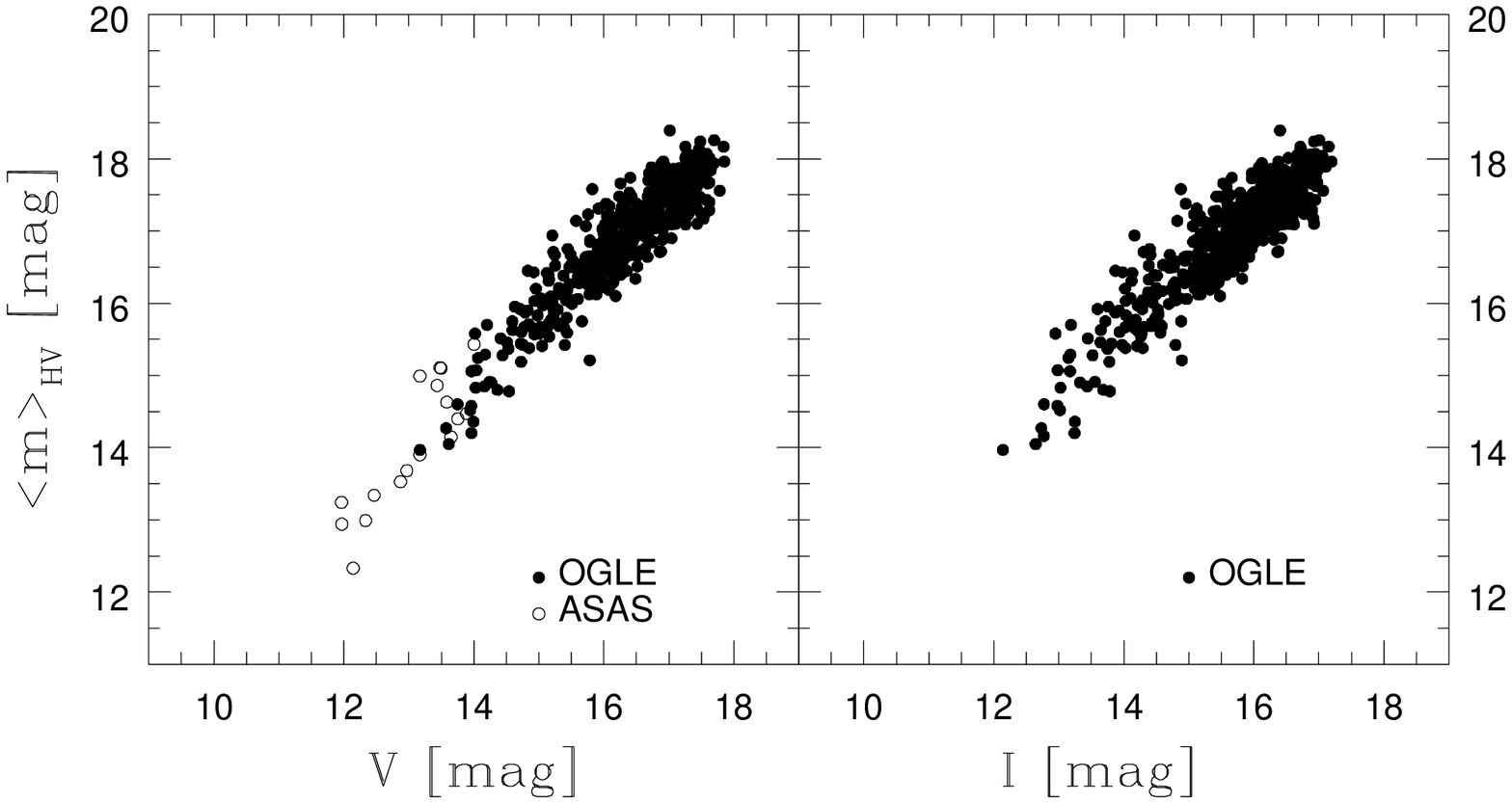,bbllx=0pt,bblly=0pt,bburx=590pt,bbury=720pt,width=12.5cm,
clip=,angle=0}
\vspace*{3pt}
\FigCap{
A comparison between the mean Harvard magnitudes and V-band magnitudes
(left panel) for 518 OGLE and seventeen ASAS Cepheids
in the Small Magellanic Cloud. Full sample of 540 OGLE Cepheids
can be presented in I band (right panel).
}
\end{figure}

\begin{figure}[htb]
\hglue-0.5cm\psfig{figure=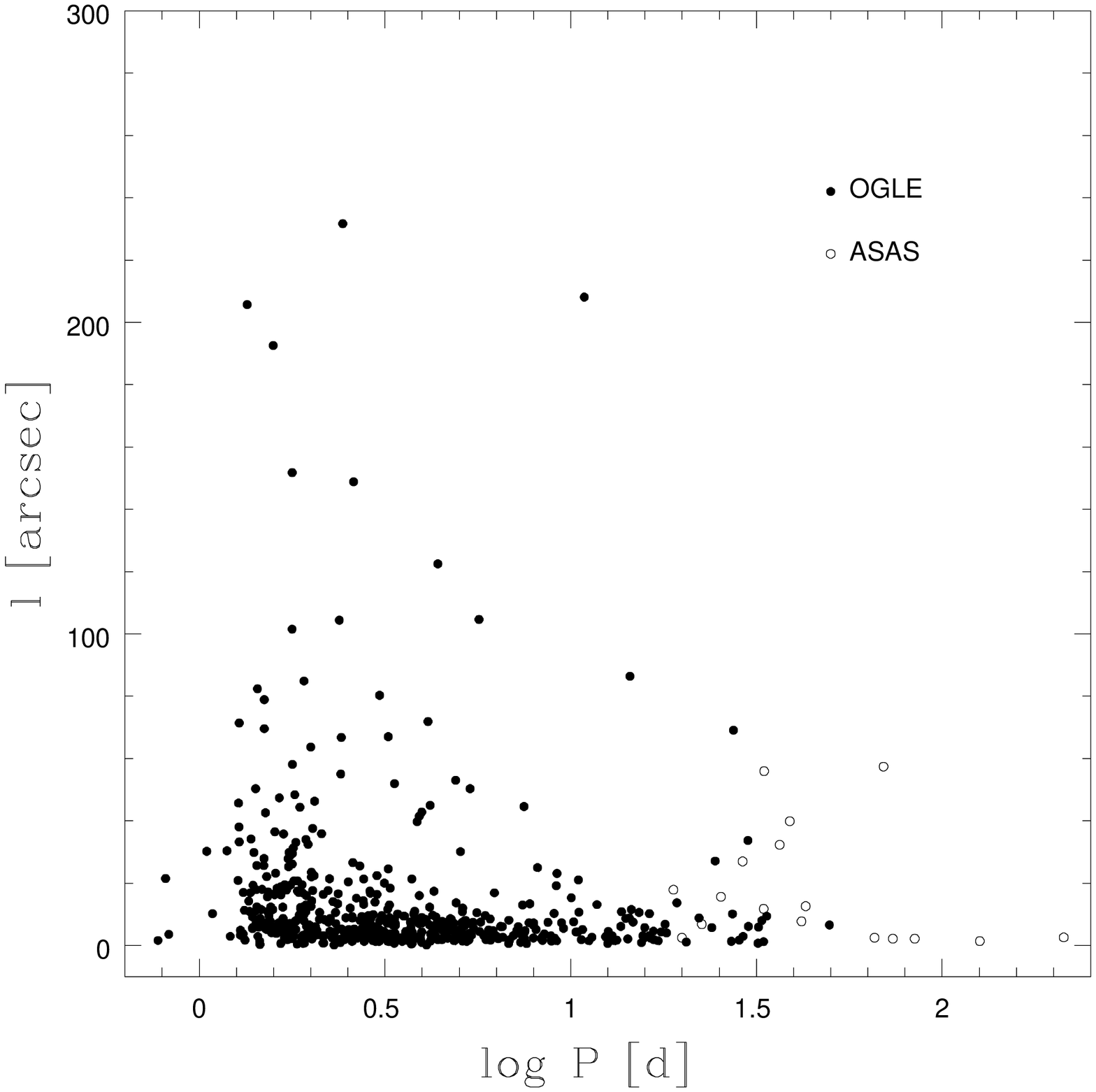,bbllx=0pt,bblly=0pt,bburx=590pt,bbury=720pt,width=12.5cm,
clip=,angle=0}
\vspace*{3pt}
\FigCap{
The difference in coordinates between the Harvard catalog and the OGLE
and ASAS catalogs is displayed as a function of period. All Cepheids
with ${\rm l > 50"}$, except two cases, have a large difference in only
one of the coordinates.
}
\end{figure}

\begin{figure}[htb]
\hglue-0.5cm\psfig{figure=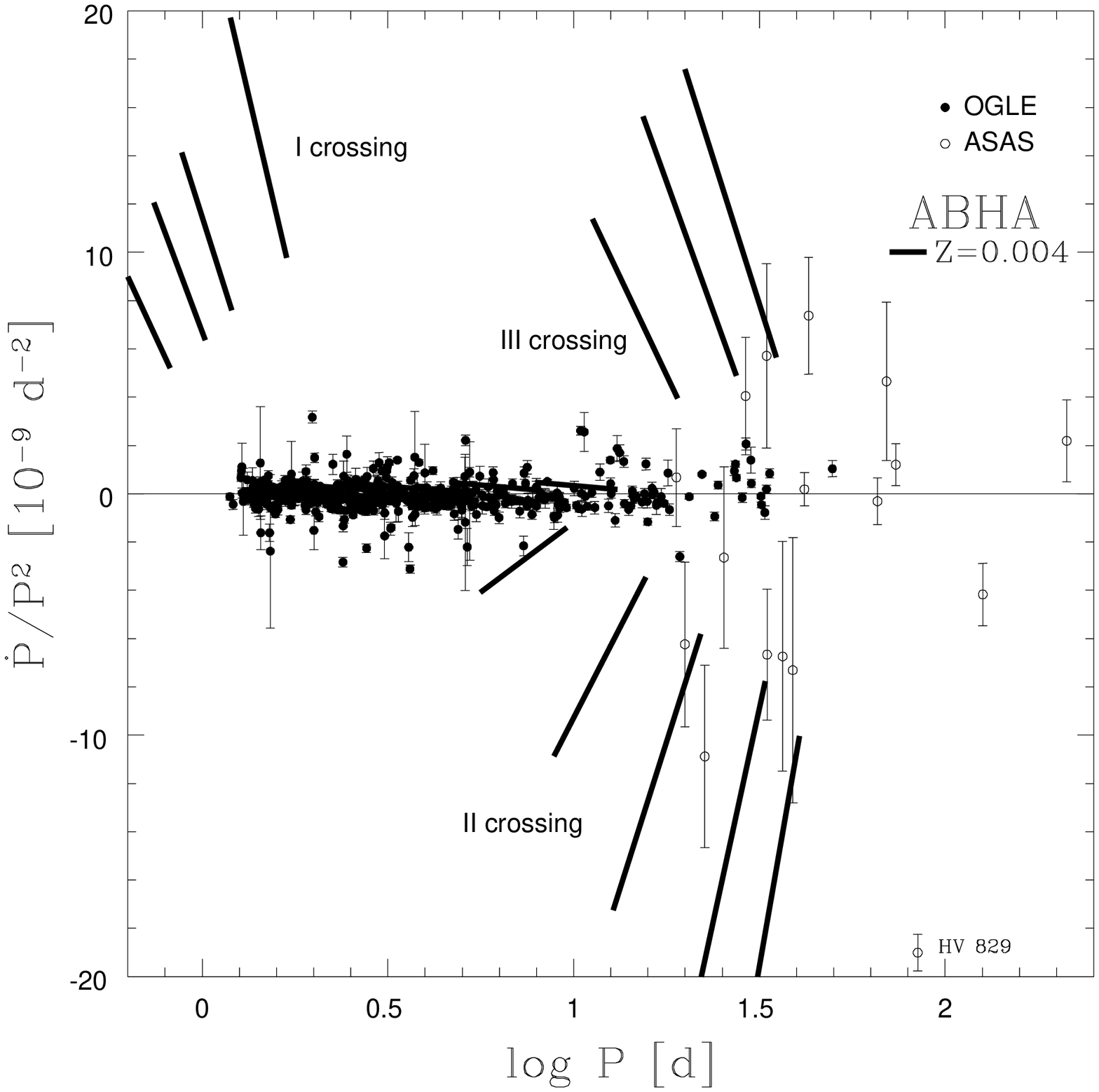,bbllx=0pt,bblly=0pt,bburx=590pt,bbury=720pt,width=12.5cm,
clip=,angle=0}
\vspace*{3pt}
\FigCap{
A comparison between the period changes predicted by ABHA (Alibert, 
Baraffe, Hauschildt, Allard 1999) and the Harvard, OGLE and ASAS
observations. The predicted values for crossings II and III
start already at $log ~ P \approx 0.1$, but they are overshadowed by
a group of observational points.
Notice none of the Cepheids on the first crossing
corresponding to the evolution on the thermal time scale.
For Cepheids with long periods there is a large disagreement.
}
\end{figure}

\begin{figure}[htb]
\hglue-0.5cm\psfig{figure=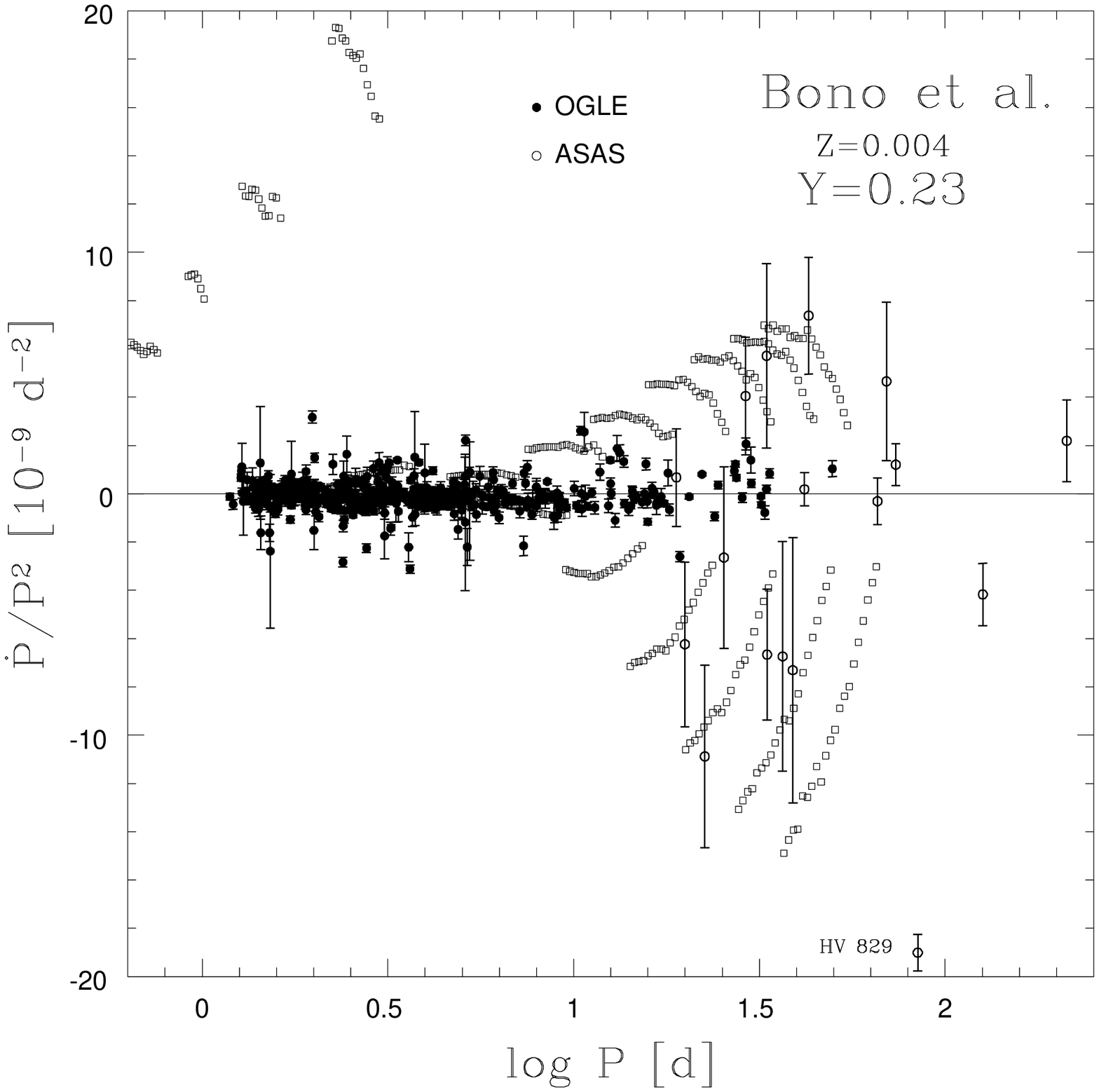,bbllx=0pt,bblly=0pt,bburx=590pt,bbury=720pt,width=12.5cm,
clip=,angle=0}
\vspace*{3pt}
\FigCap{
Period changes from the Harvard, OGLE and ASAS observations are  
compared with a set of models generated by Bono et al. (2000).
The predicted values for crossings II and III start already at
$log ~ P = 0.1$, but they are overshadowed by a group of
observational points. In general,
long period Cepheids change their periods
slower than it is predicted. Notice a relatively large
value for HV 829. This star decreased its period constantly
from 89.9 to 84.4 days in 111 years.
}

\end{figure}

\begin{figure}[htb]
\hglue-0.5cm\psfig{figure=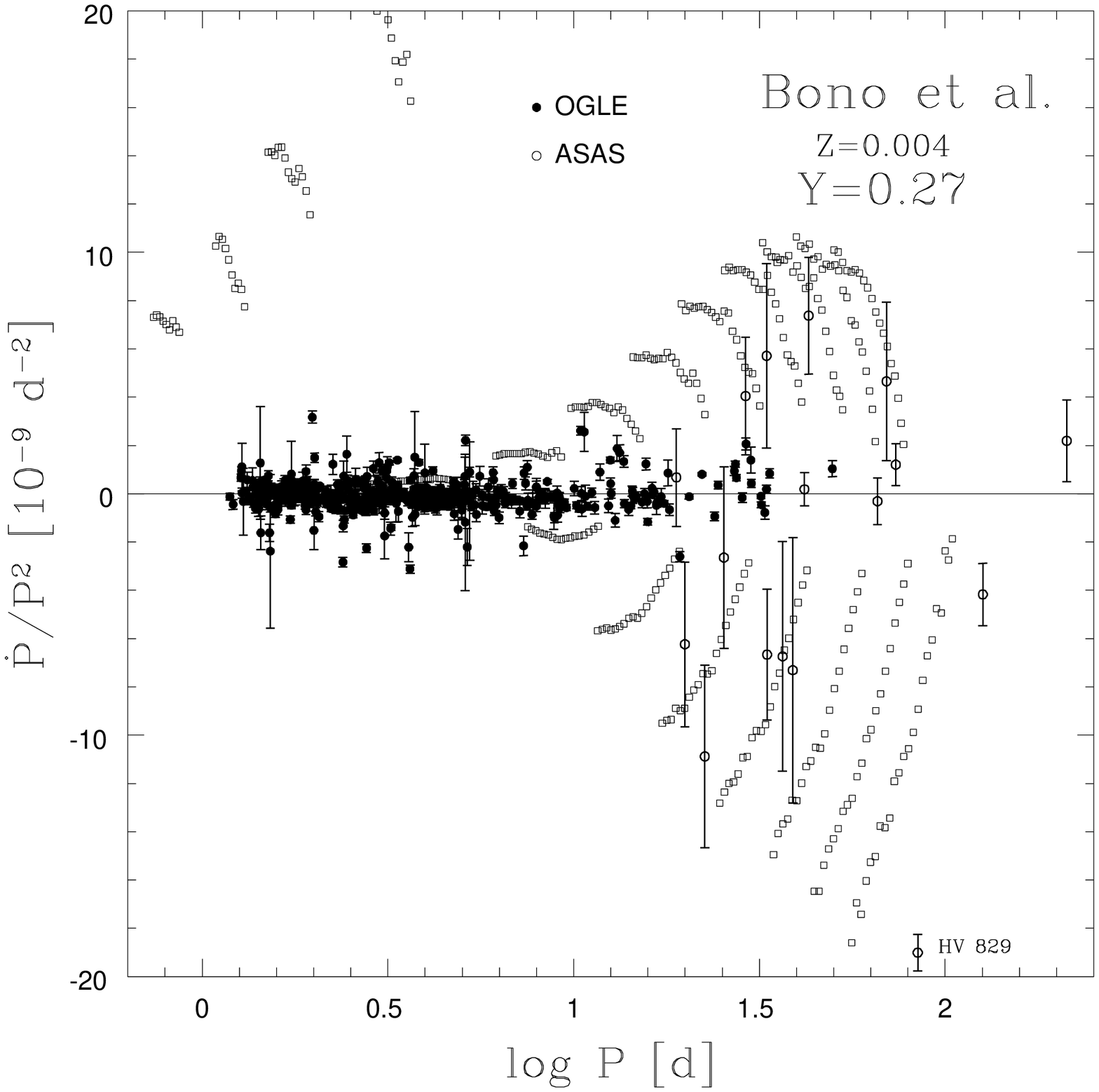,bbllx=0pt,bblly=0pt,bburx=590pt,bbury=720pt,width=12.5cm,
clip=,angle=0}
\vspace*{3pt}
\FigCap{
Period changes from the Harvard, OGLE and ASAS observations are
compared with other set of models generated by Bono et al. (2000).
A larger helium content than presented in Fig. 4 gives slightly
larger range of period changes for crossing III.
}
\end{figure}

\begin{figure}[htb]
\hglue-0.5cm\psfig{figure=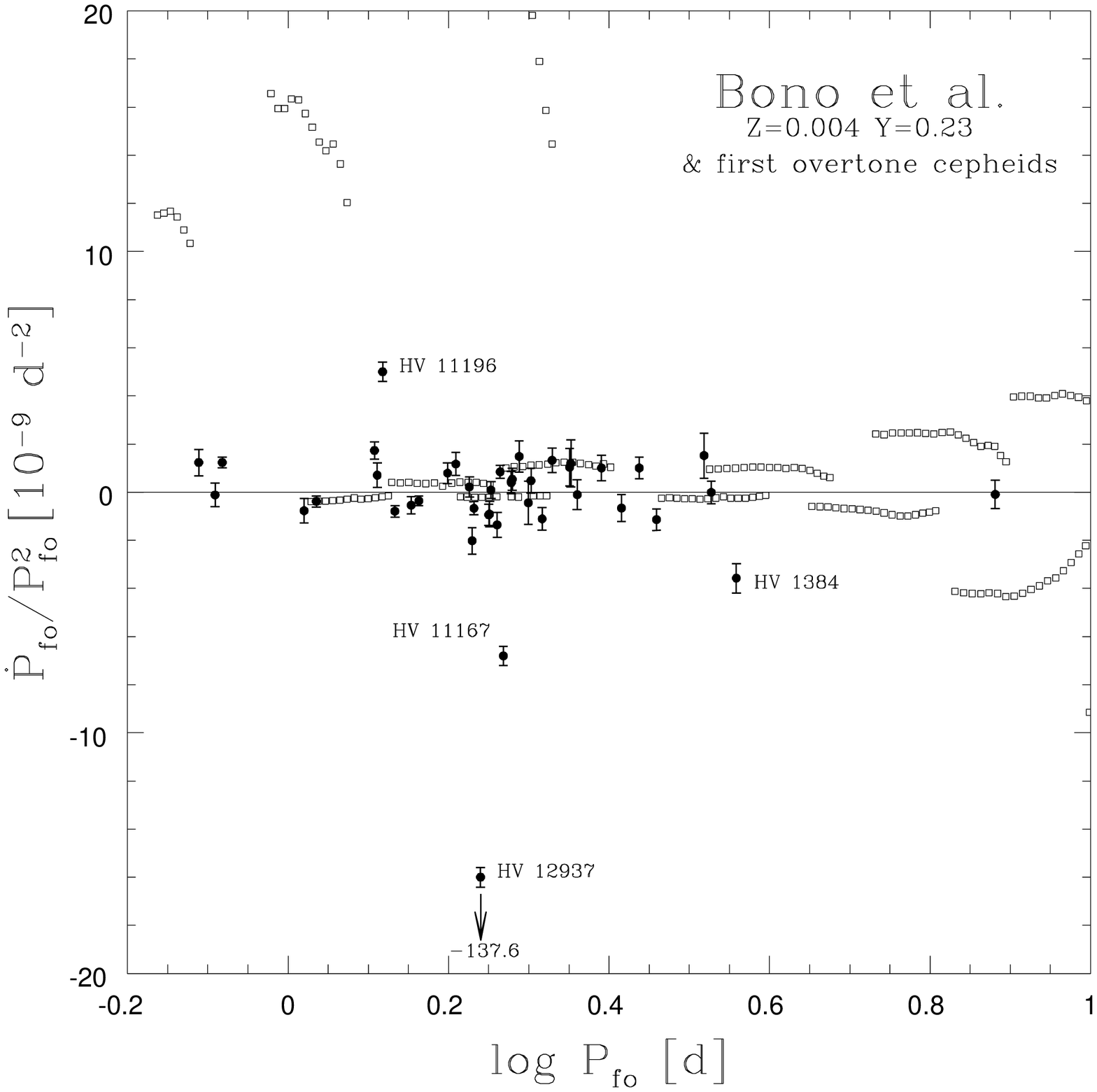,bbllx=0pt,bblly=0pt,bburx=590pt,bbury=720pt,width=12.5cm,
clip=,angle=0}
\vspace*{3pt}
\FigCap{
Comparison between the predicted and observed period changes
for 42 first overtone Cepheids. HV 12937 appears to have a strange
rate of period change.
}
\end{figure}

\begin{figure}[htb]
\hglue-0.5cm\psfig{figure=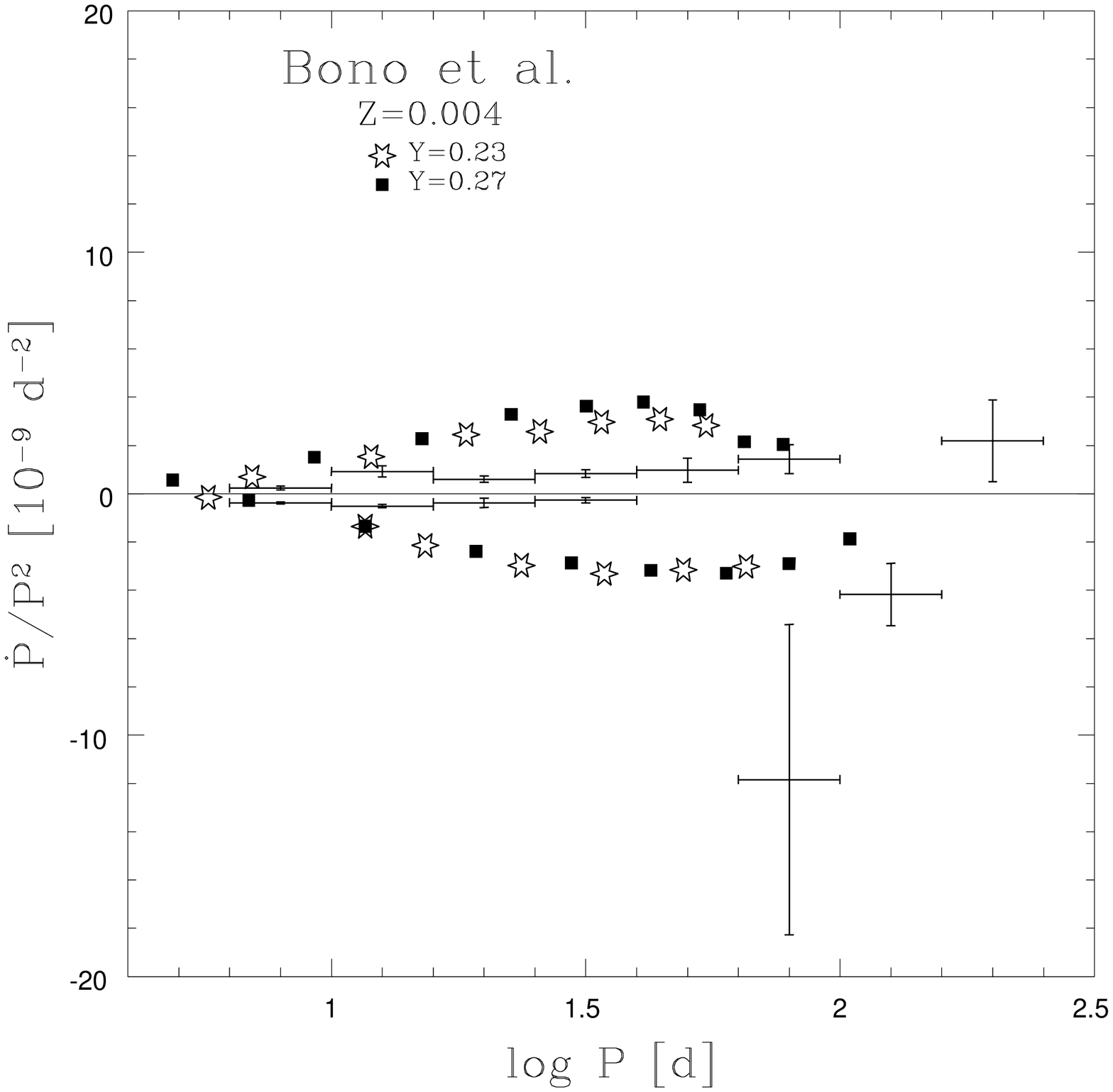,bbllx=0pt,bblly=0pt,bburx=590pt,bbury=720pt,width=12.5cm,
clip=,angle=0}
\vspace*{3pt}
\FigCap{ 
Comparison between the predicted lowest values of period
changes for crossings II and III and the observed changes
for Cepheids with long periods.
The crosses represent weighted averages of observed
period changes calculated in ${\rm log ~ P = 0.2 }$ intervals.
Note that the models generally give a few times larger
values of the changes.
}
\end{figure}

\end{document}